\pdfoutput=1
\documentclass[11pt]{article}

\usepackage[table]{xcolor}
\usepackage{booktabs}
\usepackage{multirow}
\usepackage{array}
\usepackage{colortbl}
\usepackage{caption}
\usepackage{graphicx}
\usepackage[final]{acl}
\usepackage{hyperref}
\usepackage{graphicx}
\usepackage{booktabs}       
\usepackage{array}          
\usepackage{amsmath, amssymb, amsfonts} 
\usepackage{multirow}       
\usepackage[table]{xcolor}  
\usepackage{color}          
\definecolor{red}{RGB}{255,0,0}   
\definecolor{green}{RGB}{0,128,0} 
\usepackage{microtype}      
\usepackage[utf8]{inputenc} 
\usepackage[T1]{fontenc}    
\usepackage{times}          
\usepackage{latexsym}       
\usepackage{tabularx}
\usepackage{algorithm}
\usepackage{algpseudocode} 
\usepackage{amsmath, amssymb} 
\usepackage{algpseudocode}
\usepackage{amsmath}
\usepackage{amsfonts}
\usepackage{amssymb}
\usepackage{algorithm}
\usepackage{amsmath}
\usepackage{CJKutf8}
\usepackage{makecell}
\usepackage[bottom]{footmisc}
\usepackage[most]{tcolorbox}
\usepackage[utf8]{inputenc}
\usepackage{booktabs}
\usepackage{multirow}
\usepackage{microtype}
\usepackage{inconsolata}
\usepackage{graphicx}

\title{Boosting Data Utilization for Multilingual Dense Retrieval}

\author{
 \textbf{Chao Huang\textsuperscript{1,2}\footnotemark[1]},
 \textbf{Fengran Mo\textsuperscript{3}\footnotemark[1]},
 \textbf{Yufeng Chen\textsuperscript{1,2}},
 \textbf{Changhao Guan\textsuperscript{1,2}},
\\
 \textbf{Zhenrui Yue\textsuperscript{4}},
 \textbf{Xinyu Wang\textsuperscript{5}},
 \textbf{Jinan Xu\textsuperscript{1,2}},
 \textbf{Kaiyu Huang\textsuperscript{1,2}\footnotemark[2]}
\\
 \textsuperscript{1}Key Laboratory of Big Data \& Artificial Intelligence in Transportation \\
 (Beijing Jiaotong University), Ministry of Education \\
 \textsuperscript{2}School of Computer Science and Technology, Beijing Jiaotong University \\
 \textsuperscript{3}University of Montreal; \textsuperscript{4}University of Illinois Urbana-Champaign; 
 \textsuperscript{5}McGill University \\
 \texttt{\{huangchao,kyhuang\}@bjtu.edu.cn; fengran.mo@umontreal.ca}
}

\begin{document}
\maketitle
\renewcommand{\thefootnote}{\fnsymbol{footnote}}
\footnotetext[1]{Equal contribution.}
\footnotetext[2]{Kaiyu Huang is the corresponding author.}

\renewcommand{\thefootnote}{\arabic{footnote}}
\setcounter{footnote}{0}
\begin{abstract}
Multilingual dense retrieval aims to retrieve relevant documents across different languages based on a unified retriever model. 
The challenge lies in aligning representations of different languages in a shared vector space.
The common practice is to fine-tune the dense retriever via contrastive learning, whose effectiveness highly relies on the quality of the negative samples and the efficacy of mini-batch data. 
Different from the existing studies that focus on developing sophisticated model architecture, we propose a method to boost data utilization for multilingual dense retrieval by obtaining high-quality hard negative samples and effective mini-batch data. 
The extensive experimental results on a multilingual retrieval benchmark, MIRACL, with 16 languages demonstrate the effectiveness of our method by outperforming several existing strong baselines.
\end{abstract}

\section{Introduction}

Multilingual dense retrieval~\cite{nie2010cross,zhang2023toward} aims to retrieve relevant documents based on dense representation across multiple languages. The objective of the task is to enable the retriever models to handle queries and documents in various languages by establishing better representations for a set of languages during model training.

However, constructing unified dense representations for multiple languages within a single model is non-trivial. The challenges come up with different languages could have unique syntactic structures, vocabularies, and nuances, making it difficult for a single retriever to align their representations in a shared vector space via fine-tuning~\cite{conneau2019unsupervised,macavaney2020teaching,asai2021one,huang2023soft,lin2023maggretriever}.
Besides, the data scarcity for low-resource languages further induces difficulty during fine-tuning, due to insufficient/imbalanced annotated relevance judgments~\cite{huang2024unsupervised,thakur2024leveraging}.

\begin{figure}
    \centering
    \includegraphics[width=1\linewidth]{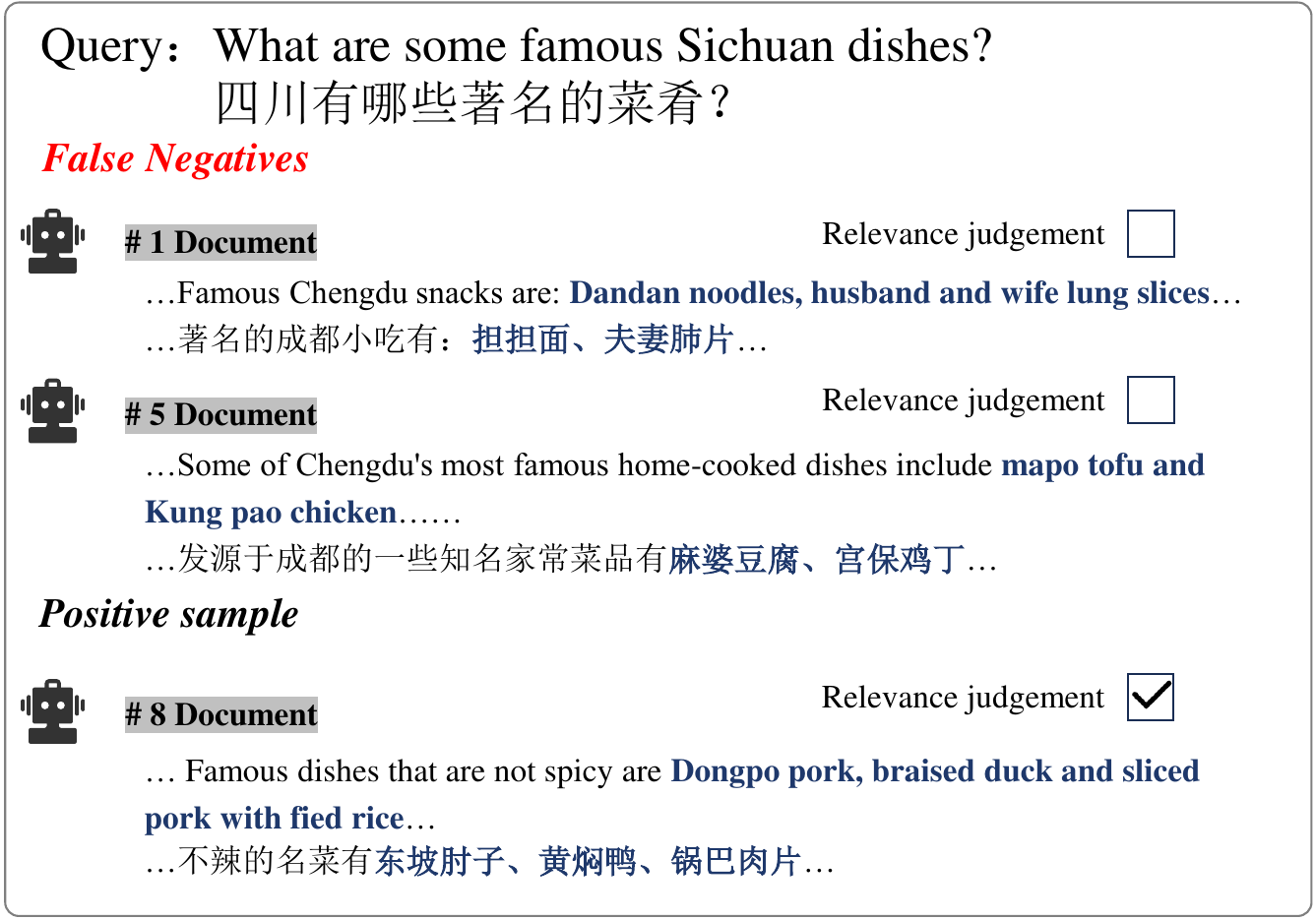}
    \caption{Example of false negatives, which refer to those that are relevant to the query but used as negatives in the ranking list due to a lack of relevance judgments.}
    \label{fig:example}
\vspace{-2ex}
\end{figure}

Existing studies attempt to address these issues from different aspects, including developing embedding models by incorporating information from multiple languages during the pre-training~\cite{devlin2018bert,izacard2021unsupervised}, fine-tuning multilingual dense retriever with additional adapter modules to allow efficient parameter sharing among models for different languages~\cite{nogueira2019multi,gururangan2020don,pfeiffer2020mad,zhan2022disentangled,lassance2023extending,hu2023language}, and integrating the multilingual pre-training and language-adapter fine-tuning~\cite{zhang2023toward}.
Most of them address the problems from the perspective of model architecture~\cite{huang2024survey,xu2025survey}.

An alternative solution is to enhance the quality of data utilization.
One common practice is constructing high-quality negative samples, which facilitates effective dense retriever fine-tuning under the contrastive learning (CL) paradigm, as demonstrated in many previous studies. 
The common way to obtain negative samples includes the BM25 negatives~\cite{karpukhin2020dense,ding2020rocketqa} and dynamic hard negatives sampling~\cite{xiong2020approximate} in ad-hoc search.
The principle is employing the top-$k$ retrieved candidate in a ranking list (e.g., pseudo relevance feedback~\cite{Xu1996QueryEU}) as hard negative candidates except for the positive ones with relevance judgment.
However, they usually introduce \textit{false negatives} -- the samples that should be relevant to the query but are not annotated, and then used as negatives for model training.
One example is shown in Figure~\ref{fig:example}, where all three documents should be considered relevant to the given query. However, since two of them do not have relevance judgment, they would be used as negative candidates.
This is a common issue since requiring the annotators to cover all existing relevance judgments is impossible. The judged part might be quite small in practice, especially on top of the huge size collection~\cite{nguyen2016ms,kwiatkowski2019natural,mo2025convmix}.
To facilitate dense retriever fine-tuning, these false negatives should be excluded when sampling the hard negative set.

In terms of multilingual retrieval, we could have more alternatives to obtain better hard negatives based on language variability and integrate them into mini-batch data, which is under-explored.
This is related to how to facilitate multilingual dense retrieval by boosting the data utilization.

Motivated by this, we design a data utilization enhanced method to improve the effectiveness of multilingual dense retrieval fine-tuning from two aspects: i) obtaining high-quality hard negatives through selection and generation, and ii) constructing effective mini-batches by adjusting language and topic semantic features.
The whole framework consists of three stages.
The first stage is to initialize the hard negative candidate set for each given query by aggregating the retrieved results from various multilingual embedding models.
Then we can select the high-quality sample by eliminating false negatives via judgments of LLMs through incorporating additional signals.
The second stage aims to inject additional hard negatives by specific LLM generation to improve the data diversity and to ensure sufficient negatives in the candidate set for sampling, i.e., the size of the negative candidate set should be equal for each query after elimination.
Finally, with the improved quality hard negative set for each query, we construct effective mini-batches by adjusting the language and topic distribution, and integrate the hard negative sampling weight determined after the data adjustment into contrastive learning to facilitate retriever fine-tuning. The query-document pairs in each mini-batch should be in the same language but have diverse topics. 
Extensive experimental results on the multilingual retrieval dataset MIRACL~\cite{zhang2023miracl} with 16 languages demonstrate the effectiveness of our method by significantly outperforming several existing strong baselines. We also provide thorough analysis experiments to understand the functionality of each stage and component.

Our contributions are summarized as follows: 
\begin{itemize}
    \item We propose a method to boost data utilization for multilingual dense retrieval fine-tuning by constructing high-quality hard negative candidates via selection and generation.
    \item  We design effective mini-batch construction strategies by adjusting the language and topic distribution among the data points, and integrate the hard negative sampling weight into contrastive learning.
    \item We demonstrate the effectiveness of our approach by outperforming several existing strong baselines on a multilingual retrieval benchmark, MIRACL.
\end{itemize}

\begin{figure*}[!t]

        \centering
    \includegraphics[width=1\linewidth]{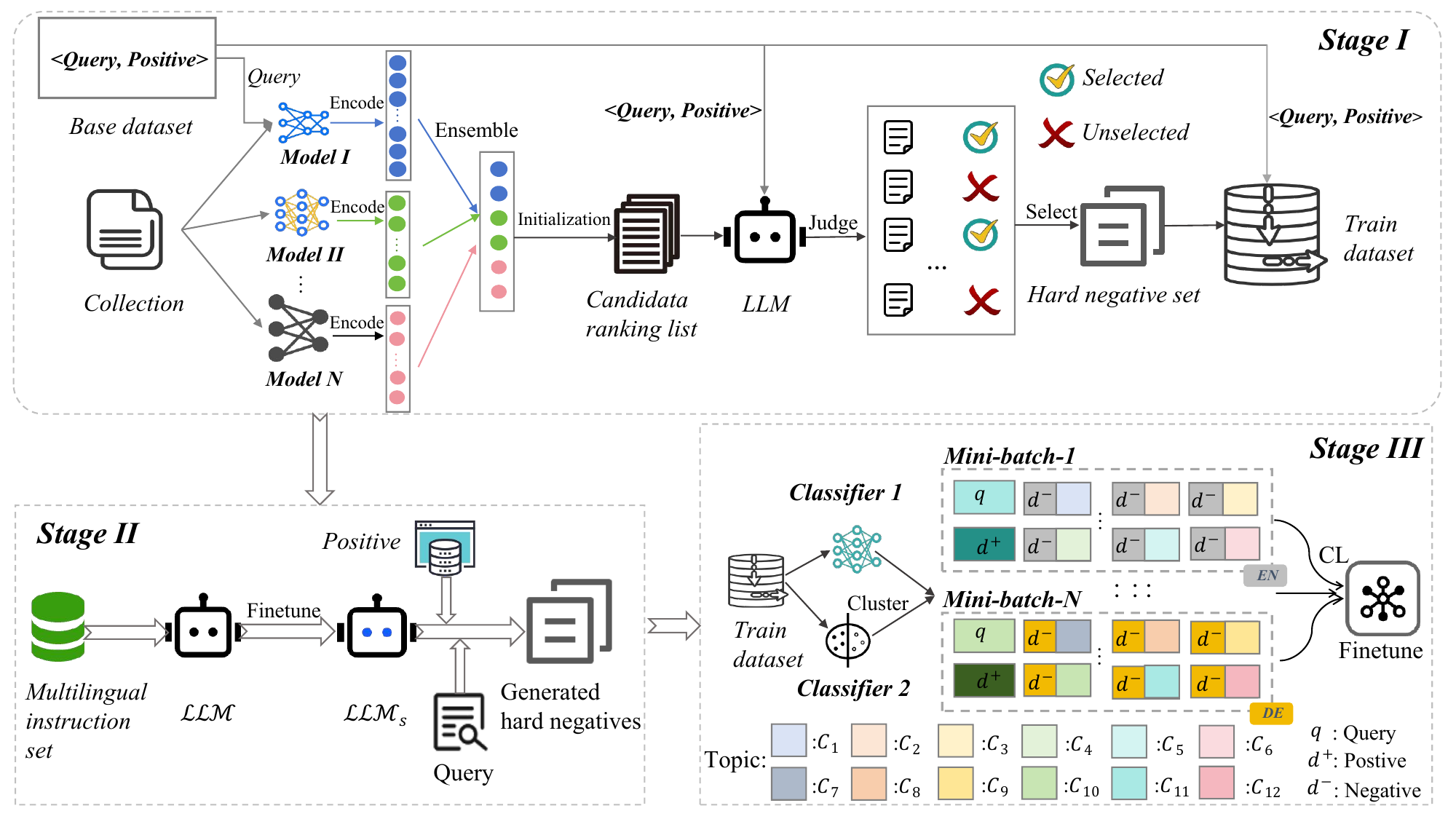}

    \caption{Overview of our framework including three stages: i) construction of hard negative set, ii) LLM-aided hard negative generation, and iii) effective mini-batch construction to facilitate contrastive learning with language and topic information adjustments, where we ensure the languages are consistent, while the topics are diverse.}
    \label{fig:overview}
\end{figure*}

\section{Related Work}
\textbf{Multilingual Information Retrieval.}
The development of multilingual information retrieval (MLIR) is important to support the demand for global information access~\cite{oard1998survey,peters2012multilingual,dwivedi2016survey}.
The advancements of MLIR can be categorized into two factors: the development of benchmarks for system evaluation by covering more languages~\cite{li2022museclir,zhang2023miracl} and the improvement of representation learning across different languages~\cite{lawrie2023neural,yang2024distillation}. 
 On top of the available resources, some studies~\cite{lin-etal-2023-maggretriever,tan2024learning,fang2024enhancing} aim to enhance zero-shot transfer capabilities through a novel dual-encoder architecture that jointly optimizes semantic alignment and lexical correspondence across languages. 
 
Besides, the other studies~\cite{multim3,xu2024self} introduce multi-granular contrastive learning combined with self-knowledge distillation to preserve language-agnostic semantic structures.
Furthermore, recent studies~\cite{ding2024data,thakur2023leveraging} explore data augmentation strategies via large language models (LLMs) and observe that synthetic multilingual training data can enhance model robustness against linguistic variations.
Different from the existing studies that focus on cross-lingual alignment, model architecture design, and data augmentation, we explore how to leverage the available data to facilitate dense retriever fine-tuning from a data utilization enhancement perspective.

\noindent \textbf{Hard Negative Mining for Dense Retrieval.}
The efficacy of hard negative in dense retrieval has been demonstrated in previous studies~\cite{karpukhin2020dense,xiong2020approximate,mo2023learning,mo2024history}.
Theoretical studies~\cite{zhou2024debiased,maharana2022curriculum,mo2024survey} demonstrate that dynamic hard negative selection mechanisms effectively enhance embedding space separability via curriculum-based sampling strategies. 
To address the scarcity of authentic hard negatives, generative models have been employed to construct contextually coherent yet semantically contradictory negative samples through controlled textual perturbation techniques~\cite{qiao2023improving,mo2023convgqr,li2024conan,su2025generating,mo2025uniconv}. 
Hybrid contrastive learning frameworks~\cite{li2022keywords,zhao2025debiased} by integrating static and dynamic negative mining strategies achieve better performance on semantic textual similarity benchmarks. 
Different from the existing methods that focus on single-language or static negative mining, our study investigates how to construct negative samples via selection and generation.

\section{Methodology}

\subsection{Task Definition}
Multilingual dense retrieval aims to retrieve relevant information across multiple languages via dense representation. 
Given a set of language $\{l\}_{t=1}^T$ and a query $q^l$, a multilingual retriever is expected to identify the relevant documents $d_q^{+}$ from the corresponding large collection $\mathcal{C}_l$ under the monolingual setting, i.e., the query $q^l$ and its candidate documents $\mathcal{D}_l=\{d_i^l\}_{i=1}^k$ are in the same language $l$, where $\mathcal{D}_l \subset \mathcal{C}_l$ and $|\mathcal{D}_l| \ll |\mathcal{C}_l|$.

\subsection{Method Overview}
Our proposed method aims to facilitate effective multilingual dense retrieval fine-tuning by constructing better hard negative samples and mini-batch data. Then, the hard negative sampling weight is integrated upon each query with the contrastive learning objective.
The method overview is presented in Figure~\ref{fig:overview}, which includes three stages: i) construction of hard negative set (Sec.~\ref{sec: Hard Negatives Set Construction}), ii) Multilingual LLM-aided hard negative generation (Sec.~\ref{sec: LLM-aided Hard Negative Generation}), and iii) effective mini-batch construction to facilitate contrastive learning (Sec.~\ref{sec: Effective Mini-Batch Construction}).

\subsection{Hard Negatives Set Construction}
\label{sec: Hard Negatives Set Construction}
The common practice to initialize a set of hard negative candidates for a specific query is to eliminate the positive sample from a top-k ranking list produced by another type of retriever, e.g., BM25~\cite{robertson2009probabilistic}. 
However, it might include a portion of false negatives due to the candidate documents at the top-rank position could still be relevant without relevance judgments according to the principle of pseudo relevant feedback~\cite{Xu1996QueryEU}. Utilizing the false negative samples as hard negative signals might be harmful for dense retriever fine-tuning.

\noindent \textbf{Hard Negative Candidate Initialization.}
To obtain the true negatives under multilingual scenarios, we first design a multilingual retriever ensemble approach for representation fusion to produce the initial candidate set.
Specifically, we employ multiple multilingual retrievers with different linguistic understanding capabilities to encode the query $q^l$ and every candidate document $d_i^l$.
Then, a feature extraction layer $\mathbf{E}$ is employed to unify the output of each encoder $f_z$ into the same dimension and concatenate them as  
$
\mathcal{V}(q^l, d_i^l) = \mathbf{E}(f_1(q^l, d_i^l)) \circ \cdots \circ \mathbf{E}(f_z(q^l, d_i^l))
$. 
Finally, the logit of the $\mathcal{V}(q^l, d_i^l)$ is used as the score to produce the top-$k$ candidate ranking list $\phi(q^l)$. 

\noindent \textbf{False Negative Selection.}
With the initial ranking list $\phi(q^l)$, we aim to identify the false negatives.
To achieve this, we leverage a large language model $\mathcal{LLM}$ to judge each candidate document $d_i^l \in \phi(q^l)$ paired with corresponding query and positive sample pair $(q^l,d_i^{+})$ via a designed prompt to produce three granularity of relevance: irrelevant, partially relevant, and highly relevant, denoted as $0$, $1$, and $2$, respectively.
Finally, only the ones judged as irrelevant remained in the hard negative candidate set $\mathcal{NC}_q$, where the size of $\mathcal{NC}_q$ is denoted as $|\mathcal{NC}_q|$.
$$
\mathcal{NC}_q = \{d_i^l \in \phi(q^l) \ | \ \mathcal{LLM}(q^l, d_{q}^{+}, d_i^l) = 0 \}
$$

\subsection{LLM-aided Hard Negative Generation}
\label{sec: LLM-aided Hard Negative Generation}
A part of the false hard negative candidates would be filtered out after the sample selection in the initial set $\mathcal{NC}_q$ for query $q^l$.
To ensure the number of samples in the hard negative candidate set $\mathcal{NC}_q$ of each query is the same for negative sampling during the multilingual dense retriever fine-tuning, we provide the supplement for the query that does not have enough negative candidates, i.e., $|\mathcal{NC}_q| < N$, where $N$ is a pre-defined constant.
To this end, we conduct specific instruction fine-tuning under a multilingual scenario for an LLM to equip it with the generation ability of hard negatives.

\noindent \textbf{Multilingual Instruction Fine-tuning.} 
The assumption is that a multilingual LLM with specific multilingual instruction fine-tuning can achieve precise negative generation with better capacity to identify multilingual samples~\cite{li2024syneg}, compared to the vanilla $\mathcal{LLM}$ used in stage one for hard negative judgment.
To implement this, we leverage the summarization-related instructions from the Alpaca dataset~\cite{taori2023stanford} as $\mathcal{I}_s$ and translate them into each language $l$ by Google Translate~\cite{googletranslate} to form a multilingual instruction set $\mathcal{I}_s^{\prime}$. 
Then, the vanilla $\mathcal{LLM}$ is fine-tuned with $\mathcal{I}_s^{\prime}$ to produce its variant $\mathcal{LLM}_s$.

\noindent \textbf{Positive-Driven Back-Forward Generation.}
With the multilingual instruction fine-tuned $\mathcal{LLM}_s$, we summarize the positive $d_q^+$ in terms of the corresponding query $q^l$, which aims to obtain query-centric key information from a lengthy document.
Then, we continue to generate a new query on top of the summary $\mathcal{S}_q^l$ and use it to obtain additional negatives $\mathcal{NC}_q^{\prime}$ via the multilingual retriever ensemble mechanism in the previous stage 1. 
We add the top candidates $\mathcal{NC}_q^{t}$ from $\mathcal{NC}_q^{\prime}$ in $\mathcal{NC}_q$ to ensure the number of final hard negatives for each query is equal to $N$, i.e., $|\mathcal{NC}_q|+|\mathcal{NC}_q^{t}| = N$.

\subsection{Effective Mini-Batch Construction to Facilitate Contrastive Learning}
\label{sec: Effective Mini-Batch Construction}
The contrastive learning (CL) paradigm is widely used in dense retriever fine-tuning due to its sophistication in leveraging the negatives.
The training objective is formulated as
$$
\mathcal{L}_{\text{MR}} = - \log  \frac{e^{\text{sim}\left(q^{l}, d^{+}_{q}\right)}} {e^{\text{sim}\left(q^{l}, d^{+}_{q}\right)} + \sum_{d^{-}_q \in \{\mathcal{D}_{-}\}} e^{\text{sim}\left(q^{l}, d^{-}_{q}\right)}}
$$
where $\text{sim}\left(q^{l}, d^{+}_{q}\right)$ denote the cosine similarity between the query and document.

For multilingual retrieval, the query $q^l$ used in the same mini-batch could be from various languages.
A natural property of CL is the usage of in-batch negatives, where each query-document pair could be negative for the other samples in the same mini-batch. Thus, increasing the difficulty of distinguishing between different samples in the same batch can increase the challenge of model training.
Following this principle, we ensure the language of each query-document pair in a mini-batch is the same, since the samples in various languages could be easier to identify due to the lower similarity, compared to those in the same language.

Besides, we also enable each mini-batch to include data samples from various topics to enhance the diversity of semantic features for fine-tuning.
The topic information is obtained by employing two classifiers ~\cite{joulin2016bag,joulin2016fasttext} to perform text classification on each positive document $d_q^+$ at different granularities.
Then, we cluster the output labels of the two classifiers corresponding to each data sample into the predefined \(C\) topics as \(C={\{C_1,C_2,\dots,C_{12}\}}\). 
Based on the language and topic information, we construct each mini-batch with monolingual and multi-topic data samples by uniform sampling, according to the number of samples for various languages and topics. This ensures that low-resource languages and underrepresented topics are adequately represented.
In addition, for each data point, we apply a weighted mechanism for negative sampling. The weight assigned to each negative is 
$$
\omega(d_q^-) = \alpha_{l}(d_q^-) + \beta_{c}(d_q^-)
$$
where $\alpha_{l}(\cdot)$ and $\beta_{c}(\cdot)$ are the language weight and topic weight based on the portion of each category.
Finally, the loss calculation for each mini-batch $\mathcal{B}$ is incorporated with the weight  and vanilla contrastive learning loss as 
$$
\mathcal{L}_{\text{final}} = \frac{1}{|\mathcal{B}|}\sum_{(q,d^+,d^-)\in \mathcal{B}} \omega(d^-)\times\mathcal{L}_{\text{MR}}
$$

\begin{table*}[!t]
    \centering
    \renewcommand{\arraystretch}{1.3}
    \resizebox{\textwidth}{!}{%
    \begin{tabular}{@{}l*{17}{c}@{}}
        \toprule
        \textbf{Model} & \textbf{Avg} & \textbf{ar} & \textbf{bn} & \textbf{en} & \textbf{es} & \textbf{fa} & \textbf{fi} & \textbf{fr} & \textbf{hi} & \textbf{id} & \textbf{ja} & \textbf{ko} & \textbf{ru} & \textbf{sw} & \textbf{te} & \textbf{th} & \textbf{zh} \\
        \midrule
        BM25 
         & 39.4 
         & 48.1 
         & 50.8 
         & 35.1 
         & 31.9 
         & 33.3 
         & 55.1 
         & 18.3 
         & 45.8 
         & 44.9 
         & 36.9 
         & 41.9 
         & 33.4 
         & 38.3 
         & 49.4 
         & 48.4 
         & 18.0 \\
        mBERT 
         & 39.9 
         & 46.2 
         & 42.8 
         & 37.6 
         & 44.8 
         & 46.2 
         & 45.9 
         & 41.8 
         & 38.6 
         & 27.8 
         & 41.4 
         & 40.1 
         & 39.8 
         & 29.2 
         & 34.4 
         & 33.2 
         & 48.9 \\ 
        mDPR 
         & 41.5 
         & 49.9 
         & 44.3 
         & 39.4 
         & 47.8 
         & 48.0 
         & 47.2 
         & 42.5 
         & 38.3 
         & 27.2 
         & 43.9 
         & 41.9 
         & 40.7 
         & 29.9 
         & 35.6 
         & 35.8 
         & 51.2 \\
        mContriever 
         & 43.3 
         & 52.5 
         & 50.1 
         & 36.4 
         & 41.8 
         & 21.5 
         & 60.2 
         & 31.4 
         & 28.6 
         & 39.2 
         & 42.4 
         & 48.3 
         & 39.1 
         & 56.0 
         & 52.8 
         & 51.7 
         & 41.0 \\
        mE5\textsubscript{large} 
         & 66.5 
         & 76.0 
         & 75.9 
         & 52.9 
         & 52.9 
         & 59.0 
         & 77.8 
         & 54.5 
         & \textbf{62.0} 
         & 52.9 
         & 70.6 
         & 66.5 
         & 67.4 
         & 74.9 
         & 84.6 
         & 80.2 
         & 56.0 \\
        BGE 
         & \underline{69.2} 
         & \underline{78.4} 
         & \underline{80.0}
         & \underline{56.9} 
         & \underline{56.1} 
         & 60.9 
         & \underline{78.6} 
         & 58.3 
         & 59.5 
         & \underline{56.1} 
         & \underline{72.8} 
         & \underline{69.9} 
         & \underline{70.1} 
         & \underline{78.7}
         & \underline{86.2} 
         & \underline{82.6} 
         & 62.7 \\
        \midrule
        Ours\textsubscript{mBERT} 
         & 57.9$^\dagger$
         & 66.2$^\dagger$
         & 65.5$^\dagger$ 
         & 48.8$^\dagger$ 
         & 48.5$^\dagger$ 
         & 52.7$^\dagger$ 
         & 64.8$^\dagger$ 
         & 53.4$^\dagger$ 
         & 45.4$^\dagger$ 
         & 44.9$^\dagger$ 
         & 61.9$^\dagger$ 
         & 56.2$^\dagger$ 
         & 55.5$^\dagger$ 
         & 64.3$^\dagger$ 
         & 76.3$^\dagger$ 
         & 67.4$^\dagger$ 
         & 54.8$^\dagger$  \\
        Ours\textsubscript{mDPR} 
         & 66.8$^\dagger$ 
         & 75.7$^\dagger$ 
         & 74.1$^\dagger$ 
         & 55.9$^\dagger$ 
         & 55.8$^\dagger$ 
         & $\underline{61.6}^{\dagger}$
         & 75.8$^\dagger$
         & \textbf{60.7}$^\dagger$
         & 53.8$^\dagger$
         & 52.1$^\dagger$ 
         & 71.2$^\dagger$ 
         & 67.8$^\dagger$
         & 65.1$^\dagger$
         & 73.9$^\dagger$ 
         & 84.2$^\dagger$ 
         & 74.8$^\dagger$
         & \textbf{65.9} \\
         Ours\textsubscript{mE5} 
         & 67.4$^\dagger$
         & 77.2$^\dagger$ 
         & 76.7$^\dagger$ 
         & 52.1 
         & 52.4 
         & 59.8$^\dagger$ 
         & 77.6 
         & 56.2$^\dagger$ 
         & \underline{61.7} 
         & 53.4$^\dagger$ 
         & 71.5$^\dagger$ 
         & 68.3$^\dagger$ 
         & 67.2 
         & 75.4$^\dagger$ 
         & 84.8$^\dagger$ 
         & 80.9$^\dagger$ 
         & 62.9$^\dagger$  \\
        Ours\textsubscript{BGE} 
         & \textbf{70.6}$^\dagger$ 
         & \textbf{80.6}$^\dagger$ 
         & \textbf{80.8}$^\dagger$ 
         & \textbf{57.6}$^\dagger$ 
         & \textbf{57.4}$^\dagger$ 
         & \textbf{62.2}$^\dagger$ 
         & \textbf{79.6}$^\dagger$ 
         & \underline{59.8}$^\dagger$ 
         & 61.4$^\dagger$ 
         & \textbf{57.5}$^\dagger$ 
         & \textbf{74.6}$^\dagger$ 
         & \textbf{71.8}$^\dagger$ 
         & \textbf{71.6}$^\dagger$ 
         & \textbf{79.6}$^\dagger$ 
         & \textbf{87.3}$^\dagger$ 
         & \textbf{83.2}$^\dagger$ 
         & \underline{65.3}$^\dagger$ \\
        \bottomrule
    \end{tabular}%
    }
    \caption{Multilingual retrieval performance with nDCG@10 score on the MIRACL dataset across 16 languages. 
    $\dagger$ denotes significant improvements with t-test at $p<0.05$ between our methods with the same corresponding backbone model. \textbf{Bold} and \underline{underline} indicate the best and the second best result.}
    \label{tab:example-tabl1}
\end{table*}

\begin{table*}[!t]
    \centering
    \renewcommand{\arraystretch}{1.3}
    \resizebox{\textwidth}{!}{%
    \begin{tabular}{@{}l l c c c c c c c c c c c c c c c c@{}}
    \toprule
    \textbf{Method} & \textbf{Avg} & \textbf{ar} & \textbf{bn} & \textbf{en} & \textbf{es} & \textbf{fa} & \textbf{fi} & \textbf{fr} & \textbf{hi} & \textbf{id} & \textbf{ja} & \textbf{ko} & \textbf{ru} & \textbf{sw} & \textbf{te} & \textbf{th} & \textbf{zh}\\
    \midrule
    Naive Top-K          & 67.6 & 78.6 & 79.3 & 51.2 & 51.0 & 59.5 & 77.8 & 56.6 & 58.2 & 50.3 & 72.2 & 69.5 & 69.3 & 78.1 & 85.8 & 82.0 & 62.8 \\
    Top-K shifted by N   & 67.8 & 78.1 & 79.6 & 52.1 & 51.6 & 60.2 & 77.6 & 57.1 & 58.4 & 51.4 & 71.8 & 69.8 & 69.4 & 77.6 & 85.4 & 82.1 & 63.2 \\
    TopK-Abs             & 67.8 & 78.4 & 79.2 & 51.9 & 51.8 & 59.8 & 77.8 & 57.3 & 58.2 & 51.8 & 71.6 & 69.9 & 69.5 & 78.2 & 85.2 & 81.8 & 63.0 \\
    TopK-MarginPos       & 67.9 & 77.8 & 79.1 & 52.8 & 52.2 & 60.2 & 78.0 & 57.5 & 58.6 & 52.2 & 71.4 & 69.7 & 68.8 & 78.5 & 85.6 & 82.2 & 62.5 \\
    TopK-PercPos         & 68.1 & 78.8 & 79.6 & 52.4 & 52.8 & 60.0 & 77.5 & 57.6 & 58.8 & 52.4 & 72.4 & 69.9 & 69.4 & 78.3 & 85.9 & 81.6 & 63.4 \\
    Ours 
                         & \textbf{70.6} & \textbf{80.6} & \textbf{80.8} & \textbf{57.6} & \textbf{57.4} & \textbf{62.2} & \textbf{79.6} & \textbf{59.8} & \textbf{61.4} & \textbf{57.5} & \textbf{74.6} & \textbf{71.8} & \textbf{70.8} & \textbf{79.6} & \textbf{87.3} & \textbf{83.2} & \textbf{65.3} \\
    \bottomrule
    \end{tabular}%
    }
    \caption{Comparison of different hard negative construction methods based on the BGE model for multilingual retrieval evaluated on the MIRACL dataset with nDCG@10 scores. \textbf{Bold} indicates the best result.}
    \label{tab:example-tabl2}
\vspace{-2ex}
\end{table*}

\section{Experiments}
\subsection{Experimental Setup}
\noindent\textbf{Datasets and Metrics.}
We evaluate our model on the multilingual retrieval benchmark MIRACL~\cite{zhang2023miracl}. Following prior study~\cite{zhang2023toward}, we conduct dense retrieval via Pyserini~\cite{lin2022pretrained} and use nDCG@10 and Recall@100 as evaluation metrics. More details are provided in Appendix~\ref{A}. 

\noindent\textbf{Implementation Details.}
We conduct multilingual dense retriever fine-tuning via mixing all languages as training data based on Tevatron~\cite{gao2022tevatron}.
The validation set is 10\% of the training set with a random sample.
Then, the trained retriever is applied to each language for evaluation.
We employ mBERT~\cite{devlin2019bert}, mDPR~\cite{zhang2023miracl}, mE5~\cite{wang2022text} and  BGE~\cite{chen2024bge} as the base models of our method. For hard negatives selection and generation, we employ the Llama-3.1-70B-instruct~\cite{dubey2024llama} model. For topic information in mini-batch construction, we use two classification models trained on the DBpedia Ontology and Yahoo Answers datasets provided in fastText model~\cite{joulin2016bag,joulin2016fasttext}. More implementation details are provided in Appendix~\ref{B1} and our released code at \url{https://github.com/miaomiao1205/xir_BDUMDR}.

\noindent\textbf{Baselines Methods.}
We compare our method with several widely-used and strong baselines, including: (1) \textbf{BM25}~\cite{robertson2009probabilistic}: an unsupervised lexical match retriever with strong generalization ability, (2) \textbf{mBERT}~\cite{devlin2019bert}: a multilingual version of BERT that provides dense contextualized representations, (3) \textbf{mDPR}~\cite{zhang2023miracl}: a dense passage retriever fine-tuned with contrastive learning on  Enlish MS MARCO~\cite{bajaj2016ms} dataset, (4) \textbf{mContriever}~\cite{izacard2021unsupervised}: an unsupervised multilingual dense retriever trained on Enlish MS MARCO version, (5)  \textbf{$\text{mE5}_{\text{large}}$}~\cite{wang2022text}: a multilingual text embedding model optimized for retrieval and semantic similarity tasks, and (6) \textbf{BGE}~\cite{chen2024bge}: a state-of-the-art multilingual embedding model designed for cross-language retrieval and semantic matching.

In addition, we also compare our hard negatives construction approach with several existing strategies, including: (1) \textbf{Naive Top-K}~\cite{karpukhin2020dense}: Using top-k retrieved candidate documents except the positive ones as hard negatives, (2) \textbf{Top-K shifted by N}~\cite{xiao2023c}: Removing top-N retrieved candidate documents first and using the remaining as hard negatives, (3) \textbf{TopK-Abs}~\cite{lee2024nv,merrick2024arctic,ding2020rocketqa}: Using top-N retrieved candidate documents whose similarity score is lower than a pre-defined threshold as hard negatives, (4) \textbf{TopK-MarginPos}~\cite{moreira2024nv}: The threshold is set as the upper-bound similarity score of the positive minus a fixed margin, and (5) \textbf{TopK-PercPos}~\cite{moreira2024nv}: The threshold is determined by the percentage of the maximum similarity score of the positive. 
For Top-K shifted by N, TopK-Abs, TopK-MarginPos, and TopK-PercPos, we set the hyperparameters as N=10, 0.6, 0.15, and 90\% for their optimal performance, respectively. More configuration details are provided in Appendix~\ref{B3}.

\subsection{Main Results}
The main results on MIRACL datasets with 16 languages are presented in Table~\ref{tab:example-tabl1}.

We observe that our method outperforms baseline methods on most languages, except slightly lower on a few high-resource ones, e.g., French (fr) and Chinese (zh). Specifically, we achieve 1.4\% absolute gain compared to the state-of-the-art BGE on average scores. The superior effectiveness can be attributed to two aspects: (1) the high-quality hard negatives mined by our multilingual retriever ensemble mechanism and the aid from multilingual LLM for hard negative generation further enhance the semantic discrimination ability, and (2) the richer supervision signals and semantic information provided by the effective mini-batch construction, which increase the challenge for the model during training. Moreover, adapting our method to different backbone models consistently yields improved performance, with BGE generally achieving the best results, except in French (fr) and Chinese (zh). Such a phenomenon indicates the feasibility of our method to further improve the multilingual retrieval performance on top of any sophisticated models.

\begin{table}[!t]
\centering
\small  
\begin{tabularx}{0.96\linewidth}{l *{2}{>{\centering\arraybackslash}X}}  
\toprule
\textbf{Ablation} & \textbf{nDCG@10} & \textbf{Recall@100} \\ 
\midrule
Full Model  & 70.6 & 95.9 \\
\hspace{2mm} w/o Stage 1 & 66.9 & 93.1 \\
\hspace{2mm} w/o Stage 2 & 68.5 & 94.3 \\
\hspace{2mm} w/o Stage 3 & 68.8 & 94.5 \\
\bottomrule
\end{tabularx}
\caption{Ablation on three stages of our methods based on the BGE on MIRACL. i) Stage 1: Multilingual Retriever Ensemble for Hard Negatives Set Construction, ii) Stage 2: LLM-aided Hard Negative Generation, and iii) Stage 3: Effective Mini-Batch Construction.}
\label{tab:example-tabl3}
\vspace{-3ex}  
\end{table}

\subsection{Comparison among Hard Negative Mining Methods}
We compare our strategy with several existing approaches based on BGE to validate our negative construction mechanism, which filters false negatives and leverages a multilingual LLM.
The results are reported in Table~\ref{tab:example-tabl2}, which shows that our strategy consistently achieves the best results on MIRACL across all languages. 
Specifically, our strategy outperforms the second-best method (TopK-PercPos) by 2.5\% absolute improvement, which demonstrates our better effectiveness.
This is contributed by the ability of our method to capture complex semantic relationships, and thus, more high-quality hard negative samples could be selected for model training.
The additional experimental results on top of other backbone models are provided in Appendix \ref{C}.

\begin{table}[!t]
\centering
\small  
\begin{tabularx}{0.96\linewidth}{l *{2}{>{\centering\arraybackslash}X}}  
\toprule
\textbf{Ablation} & \textbf{nDCG@10} & \textbf{Recall@100} \\ 
\midrule
Full Model  & 70.6 & 95.9 \\
\hspace{2mm} w/o mE5\textsubscript{large} & 69.2 & 95.0 \\
\hspace{2mm} w/o BGE & 69.1 & 94.9 \\
\bottomrule
\end{tabularx}
\caption{Ablation on hard negative samples initialization by integrating various models with average scores on MIRACL.}
\label{tab:example-tabl4}
\vspace{-2ex}  
\end{table}

\subsection{More Comparison}
\paragraph{Ablation Study.} We investigate the effectiveness of each component within our methods on the MIRACL.
The results are shown in Table~\ref{tab:example-tabl3}.
We observe that each component can contribute about 2\% absolute gain of the NDCG@10 score, and the effectiveness of stage 1 is more obvious than the other two.
Such results indicate the importance of maintaining high-quality samples in the hard negatives candidate set, which is consistent with previous studies~\cite{karpukhin2020dense,zhang2023miracl}. Then, continuing to polish the candidate set (e.g., generating new samples from LLM) can further enhance the utilization of hard negatives for contrastive retriever fine-tuning.
Additionally, we can also observe that the model performs worse than BGE without combining all three stages. This is because the functionality of our three stages is consistent, including data quality detection, data sample generation, and diversity utilization in mini-batches. Thus, it is possible that only combining them to obtain optimal results, especially when the backbone model is powerful, e.g., the training procedure of BGE, might already integrate some of these data utilization aspects.

\begin{table}[!t]
\centering
\small  
\begin{tabularx}{0.96\linewidth}{l *{2}{>{\centering\arraybackslash}X}}  
\toprule
\textbf{Ablation} & \textbf{nDCG@10} & \textbf{Recall@100} \\ 
\midrule
Full Model  & 70.6 & 95.9 \\
\hspace{2mm} w/o LLM judgment & 67.5 & 93.3 \\
\hspace{2mm} w/o ground-truth & 68.1 & 93.8 \\
\bottomrule
\end{tabularx}
\caption{Ablation on hard negative samples filtering by providing various references with average scores.}
\label{tab:example-tabl5}
\vspace{-2ex}  
\end{table}

\paragraph{Impact of Hard Negatives Set Construction.}

We investigate the impact of hard negative candidate initialization and false negative filtering in our hard negative construction mechanism.
Table~\ref{tab:example-tabl4} shows the performance for the initialization with different models.
We observe a performance drop when removing either model's retrieved results for the integration on the construction of candidate hard negatives, i.e., without mE5\textsubscript{large} or BGE, which suggests that relying on a single model is insufficient for generating high-quality hard negative candidates.

For false negative filtering, Table~\ref{tab:example-tabl5} presents the score with different filtering approaches.
We find that either removing the ground-truth information or the LLM judgment results in a performance drop. Besides, the results indicate that the LLM judgment contributes more to identifying false negatives, which confirms our conjecture that the multilingual ability of LLMs is beneficial to select hard negatives in multilingual dense retrieval.

\paragraph{Impact of the Strategies for LLM-aided Hard Negative Generation.}

In addition to identifying false negatives, we also utilize LLMs for hard negative generation.
Table~\ref{tab:example-tabl6} shows the impact of using various mechanisms for the negative generation.
We can observe that both multilingual instruction fine-tuning (MIFT) and positive-driven back-forward generation (PDBG) can improve the retrieval performance. The improvement can be attributed to MIFT, which directly impacts the LLM's understanding of task requirements, thereby enabling PDBG to generate additional hard negatives in a complementary paradigm that facilitates retriever fine-tuning.

\begin{table}[!t]
\centering
\small  
\begin{tabularx}{0.96\linewidth}{l *{2}{>{\centering\arraybackslash}X}}  
\toprule
\textbf{Ablation} & \textbf{nDCG@10} & \textbf{Recall@100} \\ 
\midrule
Full Model  & 70.6 & 95.9 \\
\hspace{2mm} w/o MIFT & 69.4 & 95.2 \\
\hspace{2mm} w/o PDBG & 69.2 & 95.1 \\
\bottomrule
\end{tabularx}
\caption{Ablation on LLM-aided hard negative generation via different strategies with average scores.}
\label{tab:example-tabl6}
\vspace{-2ex}  
\end{table}

\begin{table}[!t]
\centering
\small  
\begin{tabularx}{0.96\linewidth}{l *{2}{>{\centering\arraybackslash}X}}  
\toprule
\textbf{Method} & \textbf{nDCG@10} & \textbf{Recall@100} \\ 
\midrule
Full Model & 70.6 & 95.9 \\
\hspace{2mm} w/o Topic Balance & 70.3 & 95.6 \\
\hspace{2mm} w/o Same Language & 69.4 & 94.7 \\
\hspace{2mm} w/o Both & 68.5 & 94.3 \\
\bottomrule
\end{tabularx}
\caption{Performance of constructed effective mini-batch for multilingual dense retriever fine-tuning.}
\label{tab:example-tabl7}
\vspace{-2ex}  
\end{table}

\begin{table*}[!t]
\centering
\resizebox{\textwidth}{!}{
\begin{tabular}{lccccccccccccccccc}
\toprule
Hard Negatives & Avg. & ar & bn & en & es & fa & fi & fr & hi & id & ja & ko & ru & sw & te & th & zh \\
\midrule
From Retrieval & 90.9 & 92.4 & 97.1 & 80.6 & 83.9 & 90.9 & 92.2 & 88.0 & 92.9 & 90.0 & 88.7 & 90.7 & 87.8 & 95.7 & 96.1 & 96 & 89.8 \\
From LLM & 9.1 & 7.6 & 2.9 & 19.4 & 16.1 & 9.1 & 7.8 & 12.0 & 7.1 & 10.0 & 11.3 & 9.3 & 12.2 & 4.3 & 3.9 & 4.0 & 10.2 \\
\midrule
Eliminatation & 19.5 & 17.3 & 9.4 & 32.6 & 19.7 & 19.8 & 18.2 & 27.3 & 17.1 & 20.9 & 22.9 & 19.7 & 24.5 & 12.0 & 12.1 & 12.1 & 21.3 \\
\bottomrule
\end{tabular}
}
\caption{The first two rows present the statistics of the hard negatives obtained from retrieval (stage 1) and generated from LLM (stage 2). The last row reports the percentage of the eliminated false hard negatives across all languages.}
\label{tab:example-tabl11}
\vspace{-2ex}
\end{table*}

\paragraph{Impact of Effective Mini-Batch Construction for Model Fine-tuning.}
We adjust the language and topic distribution in the mini-batch for model fine-tuning.
Table~\ref{tab:example-tabl7} reflects the effectiveness of our strategies for constructing mini-batches.
We can see that both keeping all the data points in the same language (w/o Topic Balance) and balancing topic distribution (w/o Same Language) within a mini-batch are helpful for the retriever model fine-tuning. In addition, applying them simultaneously can further improve the final performance.
These results emphasize the effectiveness of language consistency and topic equilibrium in mini-batch construction.

\begin{table}[!t]
\centering
\small  
\begin{tabular}{ccc c}
\toprule
\multicolumn{3}{c}{\textbf{Sampling Weight Type}} & \multirow{2}{*}{\textbf{nDCG@10}} \\
\cmidrule(lr){1-3}
$\omega$ & $\alpha$ & $\beta$ \\
\midrule
0.95 & 0.55 & 0.4 & 69.5 \\
0.85 & 0.45 & 0.4 & \textbf{70.6} \\
0.75 & 0.55 & 0.2 & 68.6 \\
0.65 & 0.45 & 0.2 & 69.9 \\
\bottomrule
\end{tabular}
\caption{Performance of our model with different hard negative sampling weights. $\omega$: hard negative sampling weight. $\alpha$: language weight. $\beta$: topic weight.}
\label{tab:weight}
\vspace{-2ex}  
\end{table}

\section{Analysis}

\subsection{Quantitative Analysis of Improved Hard Negatives}
We conduct a quantitative analysis to comprehensively understand the aspects of our method to improve hard negatives.
The results are shown in Table~\ref{tab:example-tabl11}.
The first row indicates the percentage of false negative samples eliminated from the initial hard negative candidates.
We can see that over 20\% samples would be filtered out, while some differences remain across various languages. The results might be related to the multilingual ability of LLMs, e.g., the high-resource languages (en, fr, ja, ru, etc.) tend to be identified much more easily. 

Since we cannot control how many false negatives would be eliminated, the generated hard negatives supplied by fine-tuned LLMs are used to enhance the diversity of hard negative candidates from another aspect. 
The corresponding statistics are shown in the second and third rows in Table~\ref{tab:example-tabl11}, where the ratio of hard negatives produced by the initialization and selection in the first stage versus the LLM-aided generated samples in the second stage is about 1:9. The better data ratio could be further explore in future study.

\subsection{Weight of Hard Negative Sampling}
During the mini-batch construction, the hard negative sampling weights are obtained automatically based on the portion of the language and topic among the data points.
To analyze the impact of hard negative sampling weights, we manually control them to conduct additional analysis. The results are shown in Table~\ref{tab:weight}.
We observe that the hard negative sampling weights could be an empirical value in terms of better retrieval performance. 
Either reducing the language weight or increasing the topic weight can improve retrieval performance.
This implies that the smaller language weight $\alpha$ urges the model to pay more attention to low-resource languages, and the higher topic weight $\beta$ provides more semantic features, both benefit for obtaining better contrastive samples.

\begin{figure}[!t]
    \centering
    \includegraphics[width=0.8\linewidth]{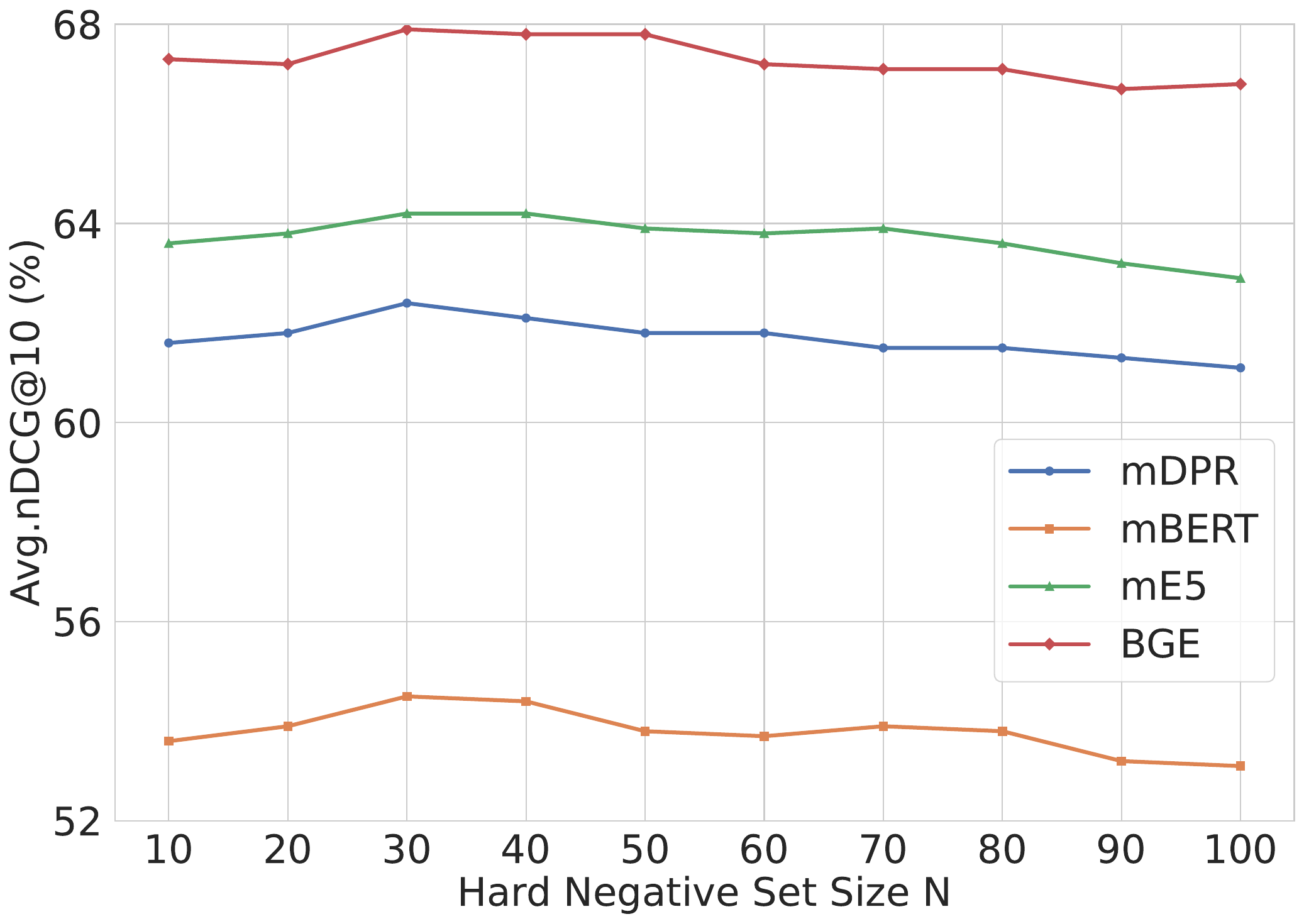}
    \caption{Model performance with average NDCG@10 on different initial hard negative candidate set sizes $N$.}
    \label{fig:size}
\vspace{-2ex}
\end{figure}

\subsection{Impact of Initial Hard Negative Candidate Set Size}
Figure~\ref{fig:size} shows the average retrieval performance on different initial hard negative candidate set sizes $N$.
We observe that for each backbone model, the performance peak usually occurs when $N$ is between 30 to 40.
Such results indicate that the optimal size for the constructed hard negative candidate set should be empirically selected, since smaller N cannot ensure sufficient potential high-quality candidates, while a larger N might increase the difficulty for identification.

\section{Conclusion}
In this work, we propose a method to boost data utilization for multilingual dense retrieval contrastive fine-tuning from two aspects: i) obtaining high-quality hard negatives through selection and generation, and ii) constructing effective mini-batches by adjusting language and topic semantic distribution.
By addressing the false hard negative issues and obtaining the high-quality ones, we integrate them into negative sampling with constructing effective mini-batches during retriever fine-tuning.
Experimental results show that our method outperforms several existing strong baselines on a multilingual retrieval benchmark and demonstrates the superior effectiveness on boosting data utilization.

\section*{Limitations}
Although with better performance, the multiple stages involved with calling LLMs in our method might raise cost concerns, which can be optimized by using an efficient or low-cost LLM. Nevertheless, it is conducted during the training phase, so no efficiency issue for inference would be raised.
Besides, the judgment of false negatives via LLMs might still be inaccurate, which is an open question in utilizing pseudo relevance feedback (PRF) in terms of information retrieval. The document in PRF could be positive or negative for a given query, while we cannot determine it absolutely without relevance judgment. 
Thus, a more sophisticated mechanism for false negative judgment can be explored in the future, and some potential human validation could be helpful to improve the accuracy of the identification mechanism. 

\section*{Acknowledgements}
The research work descried in this paper has been supported by the Fundamental Research Funds for the Central Universities (2024JBZY019) and the National Nature Science Foundation of China (No. 62476023, 62406018, 62376019). The work is also supported by the Henan Provincial Science and Technology Research Project (No. 252102210102).
The authors would like to thank the anonymous reviewers for their valuable comments and suggestions to improve this paper.

\bibliography{main}

\clearpage

\appendix

\section*{Appendix}
\addcontentsline{toc}{section}{Appendix A}  

\setcounter{section}{0}  

\section{Dataset Statistic}
\label{A}

The statistic of the MIRACL dataset~\cite{zhang2023miracl} is shown in Table \ref{tab:fulu1}. MIRACL is a multilingual retrieval benchmark dataset covering 18 languages, with 726k manual relevance judgments and a collection size of over 100 million documents. Each query is provided with an average of 10 manually verified relevance labels. 
We use 16 languages from MIRACL, except for German (de) and Yoruba (yo), due to the lack of training data. For LLM multilingual instruction fine-tuning for negative generation, we use the Alpaca dataset~\cite{taori2023stanford}.

\begin{table*}[!t]
\centering
\scriptsize
\renewcommand{\arraystretch}{1.3}
\begin{tabular}{llllllllllllll}
\toprule
\textbf{Lang} & & \textbf{All} & \textbf{Arabic} & \textbf{Bengali} & \textbf{English} & \textbf{Spanish} & \textbf{Persian} & \textbf{Finnish} & \textbf{French} & \textbf{Hindi} \\
\midrule
\textbf{ISO} & & & \textbf{ar} & \textbf{bn} & \textbf{en} & \textbf{es} & \textbf{fa} & \textbf{fi} & \textbf{fr} & \textbf{hi} \\
\midrule
\multirow{2}{*}{\textbf{Train}} & \#Q & 40,203 & 3,495 & 1,631 & 2,863 & 2,162 & 2,107 & 2,897 & 1,143 & 1,169 \\
 & \#J & 343,177 & 25,382 & 16,754 & 29,416 & 21,531 & 21,844 & 20,350 & 11,426 & 11,668 \\
\midrule
\multirow{2}{*}{\textbf{Test}} & \#Q & 13,071 & 2,896 & 411 & 799 & 648 & 632 & 1,271 & 343 & 350 \\
 & \#J & 126,076 & 29,197 & 4,206 & 8,350 & 6,443 & 6,571 & 12,008 & 3,429 & 3,494 \\
\midrule
\textbf{Passages} & & 90,416,887 & 2,061,414 & 297,265 & 32,893,221 & 10,373,953 & 2,207,172 & 1,883,509 & 14,636,953 & 506,264 \\
\midrule
\midrule
\textbf{Lang} & & \textbf{All} & \textbf{Indonesian} & \textbf{Japanese} & \textbf{Korean} & \textbf{Russian} & \textbf{Swahili} & \textbf{Telugu} & \textbf{Thai} & \textbf{Chinese} \\
\midrule
\textbf{ISO} & & & \textbf{id} & \textbf{ja} & \textbf{ko} & \textbf{ru} & \textbf{sw} & \textbf{te} & \textbf{th} & \textbf{zh} \\
\midrule
\multirow{2}{*}{\textbf{Train}} & \#Q & 40,203 & 4,071 & 3,477 & 868 & 4,683 & 1,901 & 3,452 & 2,972 & 1,312 \\
 & \#J & 343,177 & 41,358 & 34,387 & 12,767 & 33,921 & 9,359 & 18,608 & 21,293 & 13,113 \\

\midrule
\multirow{2}{*}{\textbf{Test}} & \#Q & 13,071 & 960 & 860 & 213 & 1,252 & 482 & 828 & 733 & 393 \\
 & \#J & 126,076 & 9,668 & 8,354 & 3,057 & 13,100 & 5,092 & 1,606 & 7,573 & 13,113 \\
\midrule
\textbf{Passages} & & 90,416,887 & 1,446,315 & 6,953,614 & 1,486,752 & 9,543,918 & 131,924 & 518,079 & 542,166 & 4,934,368 \\
 
\bottomrule
\end{tabular}
\caption{Statistics of MIRACL. The \#\textbf{Q} and \#\textbf{J} denote to number of queries and relevance judgments, respectively.}
\label{tab:fulu1}
\end{table*}

\begin{table*}[!t]
\centering
\small
\renewcommand{\arraystretch}{1.3}
\begin{tabular}{lll}
\toprule
\textbf{Yahoo Answers} & \textbf{DBpedia Ontology} & \textbf{MIRACL} \\
\midrule
Society \& Culture & Company & Books \& Literature (BL) \\
Science \& Mathematics & Educational Institution & Science \& Mathematics (SM) \\
Health & Artist & Life \& Health (LH) \\
Education \& Reference & Athlete & Jobs \& Education (JE) \\
Computers \& Internet & Office Holder & Computers \& Internet (CI) \\
Sports & Mean Of Transportation & Sports (SP) \\
Business \& Finance & Building & Business \& Finance (BF) \\
Entertainment \& Music & Natural Place & Politics \& Government (PG) \\
Family \& Relationships & Village & Traffic \& Transportation (TT) \\
Politics \& Government & Animal & Arts \& Entertainment (AE) \\
 & Plant & Geography (GE) \\
 & Album & Others (OT) \\
 & Film & \\
 & Written Work & \\
\bottomrule
\end{tabular}
\caption{Topic labels categorization for Yahoo Answers, DBpedia Ontology, and MIRACL datasets.}
\label{tab:fulu2}
\end{table*}

\begin{table}[ht]
\centering
\small  
\begin{tabularx}{0.96\linewidth}{l *{3}{>{\centering\arraybackslash}X}}  
\toprule
\multirow{2}{*}{{Subset}} & \multicolumn{3}{c}{Agreement with Human Judgment} \\
\cmidrule(lr){2-4}
& 0 & 1 & 2 \\
\midrule
en & 94\% & 3\% & 3\% \\
es & 91\% & 5\% & 4\% \\
zh & 92\% & 5\% & 3\% \\
hi & 87\% & 6\% & 7\% \\
bn & 90\% & 6\% & 4\% \\
\bottomrule
\end{tabularx}
\caption{The agreement between human validation and identification from LLMs on the false negative.}
\label{tab:fulu4}
\vspace{-2ex}  
\end{table}

\begin{table*}[!t]
\centering
\small
\renewcommand{\arraystretch}{1.3}
\begin{tabular}{lrrrrrrrrrrrrr}
\toprule
\textbf{Lang} & \textbf{BL} & \textbf{SM} & \textbf{LH} & \textbf{JE} & \textbf{CI} & \textbf{SP} & \textbf{BF} & \textbf{PG} & \textbf{TT} & \textbf{AE} & \textbf{GE} & \textbf{OT} & \textbf{Total} \\
\midrule
ar & 139 & 315 & 358 & 49 & 62 & 111 & 61 & 158 & 98 & 273 & 1519 & 352 & 3,497 \\
bn & 94 & 109 & 82 & 46 & 67 & 84 & 23 & 75 & 22 & 126 & 795 & 108 & 1,631 \\
en & 88 & 229 & 261 & 55 & 88 & 180 & 62 & 108 & 80 & 524 & 929 & 259 & 2,863 \\
es & 72 & 168 & 141 & 36 & 73 & 85 & 61 & 74 & 63 & 341 & 781 & 267 & 2,162 \\
fa & 101 & 204 & 180 & 55 & 99 & 87 & 65 & 59 & 65 & 304 & 688 & 200 & 2,107 \\
fi & 111 & 217 & 176 & 35 & 80 & 195 & 77 & 90 & 132 & 443 & 1011 & 330 & 2,997 \\
fr & 27 & 103 & 58 & 27 & 31 & 75 & 15 & 42 & 35 & 174 & 462 & 94 & 1,143 \\
hi & 31 & 155 & 91 & 14 & 29 & 55 & 45 & 52 & 30 & 93 & 457 & 117 & 1,169 \\
id & 122 & 378 & 160 & 59 & 223 & 89 & 153 & 152 & 113 & 502 & 1822 & 293 & 4,066 \\
ja & 146 & 184 & 150 & 74 & 96 & 286 & 107 & 127 & 145 & 626 & 1243 & 292 & 3,476 \\
ko & 12 & 112 & 39 & 6 & 26 & 16 & 21 & 50 & 20 & 59 & 437 & 70 & 868 \\
ru & 270 & 242 & 192 & 77 & 66 & 183 & 90 & 188 & 217 & 728 & 1862 & 568 & 4,683 \\
sw & 12 & 164 & 215 & 32 & 25 & 113 & 21 & 119 & 59 & 221 & 789 & 131 & 1,901 \\
te & 90 & 177 & 130 & 208 & 28 & 63 & 40 & 77 & 100 & 378 & 1935 & 226 & 3,452 \\
th & 111 & 192 & 180 & 159 & 91 & 113 & 50 & 114 & 61 & 521 & 1151 & 229 & 3,002 \\
zh & 50 & 130 & 44 & 50 & 45 & 60 & 43 & 62 & 40 & 183 & 534 & 71 & 1,312 \\
\midrule
\textbf{Total} & 1,476 & 3,079 & 2,457 & 982 & 1,129 & 1,795 & 934 & 1,547 & 1,280 & 5,496 & 16,415 & 3,607 & 40,197 \\
\bottomrule
\end{tabular}
\caption{Topic distribution results of MIRACL training set for each language.}
\label{tab:fulu3}
\end{table*}

\begin{table*}[!t]
    \centering
    \renewcommand{\arraystretch}{1.3}
    \resizebox{\textwidth}{!}{%
    \begin{tabular}{@{}l*{17}{c}@{}}
        \toprule
        \textbf{Model} & \textbf{Avg.} & \textbf{ar} & \textbf{bn} & \textbf{en} & \textbf{es} & \textbf{fa} & \textbf{fi} & \textbf{fr} & \textbf{hi} & \textbf{id} & \textbf{ja} & \textbf{ko} & \textbf{ru} & \textbf{sw} & \textbf{te} & \textbf{th} & \textbf{zh} \\
        \midrule
        \cmidrule(lr){1-18}
        BM25 
         & 78.7 
         & 88.9 
         & 90.9 
         & 81.9 
         & 70.2 
         & 73.1 
         & 89.1 
         & 65.3 
         & 86.8 
         & 90.4 
         & 80.5 
         & 78.3 
         & 66.1 
         & 70.1 
         & 83.1 
         & 88.7 
         & 56.0 \\
        mBERT 
         & 77.8 
         & 82.6 
         & 81.2 
         & 75.7 
         & 84.8 
         & 88.4 
         & 77.6 
         & 90.7 
         & 78.2 
         & 58.4 
         & 83.7 
         & 72.4 
         & 78.8 
         & 60.6 
         & 75.6 
         & 65.4 
         & 91.3 \\
        mDPR 
         & 78.8 
         & 84.1 
         & 81.9 
         & 76.8 
         & 86.4 
         & 89.8 
         & 78.8 
         & 91.5 
         & 77.6 
         & 57.3 
         & 82.5 
         & 73.7 
         & 79.7 
         & 61.6 
         & 76.2 
         & 67.8 
         & 94.4 \\
        mContriever 
         & 85.5 
         & 92.5 
         & 92.1 
         & 79.7 
         & 84.1 
         & 65.4 
         & 95.3 
         & 82.4 
         & 64.6 
         & 80.2 
         & 87.8 
         & 87.5 
         & 85.0 
         & 91.1 
         & 96.1 
         & 93.6 
         & 90.3 \\
        mE5\textsubscript{large} 
         & 94.4 
         & 97.3 
         & 98.2 
         & 87.6 
         & 89.1 
         & 92.9 
         & \textbf{98.1} 
         & 90.6 
         & 93.9 
         & 87.9 
         & 97.1 
         & 93.4 
         & 95.5 
         & 96.7 
         & 99.2 
         & 98.9 
         & 93.3 \\
        BGE 
         & \underline{95.6} 
         & \underline{97.6} 
         & \underline{98.7} 
         & \textbf{90.7} 
         & \textbf{91.1} 
         & \underline{94.0} 
         & 97.9 
         & 93.8 
         & \textbf{94.4} 
         & \textbf{90.5} 
         & \underline{97.5} 
         & \underline{95.5} 
         & \underline{95.9} 
         & \textbf{97.2} 
         & \textbf{99.4} 
         & 99.1 
         & 96.9 \\
        \midrule
        
        Ours\textsubscript{mBERT} 
         & 87.8$^\dagger$ 
         & 91.2$^\dagger$
         & 92.1$^\dagger$ 
         & 83.4$^\dagger$ 
         & 83.2 
         & 87.8 
         & 90.8$^\dagger$ 
         & 90.9$^\dagger$ 
         & 81.6$^\dagger$ 
         & 77.6$^\dagger$ 
         & 91.1$^\dagger$ 
         & 84.3$^\dagger$ 
         & 86.2$^\dagger$ 
         & 90.6$^\dagger$ 
         & 94.9$^\dagger$ 
         & 89.2$^\dagger$ 
         & 90.8 \\
        Ours\textsubscript{mDPR} 
         & 94.4$^\dagger$ 
         & 96.8$^\dagger$ 
         & 96.3$^\dagger$ 
         & $\underline{90.1}^{\dagger}$
         & $\underline{90.8}^{\dagger}$
         & 93.9$^\dagger$ 
         & 96.6$^\dagger$ 
         & \textbf{96.2}$^\dagger$ 
         & 90.2$^\dagger$ 
         & 86.8$^\dagger$ 
         & 96.6$^\dagger$ 
         & 94.8$^\dagger$ 
         & 94.5$^\dagger$ 
         & 94.8$^\dagger$ 
         & 97.4$^\dagger$ 
         & 96.6$^\dagger$ 
         & \underline{97.6} \\
        Ours\textsubscript{mE5} 
         & 94.8$^\dagger$ 
         & 97.4$^\dagger$ 
         & 98.4$^\dagger$ 
         & 86.9 
         & 88.6 
         & 93.4$^\dagger$ 
         & \underline{97.9} 
         & 92.2$^\dagger$ 
         & 93.4 
         & 88.4$^\dagger$ 
         & 97.3$^\dagger$ 
         & 94.4$^\dagger$ 
         & 95.4 
         & 96.2 
         & 99.3$^\dagger$ 
         & \textbf{99.2}$^\dagger$ 
         & \textbf{98.6}$^\dagger$ \\
        Ours\textsubscript{BGE} 
         & \textbf{95.9}$^\dagger$ 
         & \textbf{97.9}$^\dagger$ 
         & \textbf{98.9}$^\dagger$ 
         & 89.9 
         & 90.6 
         & \textbf{94.9}$^\dagger$ 
         & 98.1$^\dagger$ 
         & $\underline{95.4}^{\dagger}$ 
         & \underline{94.3} 
         & \underline{90.4} 
         & \textbf{98.1}$^\dagger$ 
         & \textbf{96.1}$^\dagger$ 
         & \textbf{96.3}$^\dagger$ 
         & \underline{97.0} 
         & \underline{99.3} 
         & \underline{99.1}  
         & 97.3$^\dagger$ \\
        \bottomrule
    \end{tabular}%
    }
    \caption{Multilingual retrieval performance with Recall@100 score on the MIRACL dataset across 16 languages. 
    $\dagger$ denotes significant improvements with t-test at $p<0.05$ between our methods with the same corresponding backbone model. \textbf{Bold} and \underline{underline} indicate the best and the second best result, respectively.}
    \label{tab:fulu5}
\end{table*}

\section{Implementation Details}
\label{B}
\subsection{Hyperparameter Setting}
\label{B1}
We implement all models by Pytorch~\cite{paszke2019pytorch} and Huggingface's Transformers library~\cite{wolf2019huggingface}. All experiments are conducted on an Nvidia A100 80G GPU. For multilingual dense retriever training, we use the Adam optimizer with maximum query and paragraph lengths set to 64 and 256, respectively, and set the batch size as 24. The learning rate is set to 3e-6 and the number of epochs to 16 for mBERT and mDPR. For mE5 and BGE, the learning rate is 2e-6 and the number of epochs is 20. Additionally, for each query, we randomly sample one positive sample and 7 hard negative samples. 

\subsection{Hard Negatives Set Construction}
\label{B2}
\noindent \textbf{False Hard Negatives Selection.}
For each hard negative sample in the candidate set, we prompt GPT-4o (2024-11-20) to select the true hard negatives as shown in Table~\ref{tab:fulu9}. The size of the hard negatives candidate set is set to 40. Besides, we use the positive sample as a reference to score each hard negative candidate from two aspects, information completeness and accuracy with three granularities -- (0,1,2).
The final scores will be their combination.
The candidate with a final score of 2 is considered a false hard negative sample. 

\noindent \textbf{LLM-aided Hard Negative Generation.}
We set the sampling size of hard negatives to 30 during the training phase for each query.
Therefore, after the filtering of false negative samples, if the number of hard negative samples is less than 30, we will deploy LLM-aided hard negative generation to supply the sample number to 30.

\subsection{Configuration of Compared Hard Negative Mining Methods}
\label{B3}

The configurations for the compared hard negative mining methods are selected according to hyperparameter tuning.
For the Top-K shifted by N method, the configuration range for N is [0, 100], with an interval of 10. For the TopK-Abs, TopK-MarginPos, and TopK-PercPos methods, the threshold/margin values range from [0, 1], with increments of 0.1 for TopK-Abs and 0.05 for TopK-MarginPos and TopK-PercPos, respectively. 
We observe that the Top-K shifted by N method performs best when N is set to 10, i.e., when the top-10 ranked negative samples are discarded.  
For TopK-Abs, TopK-MarginPos, and TopK-PercPos approaches, we find that the model achieves the best performance when the threshold is set to 0.6, 0.15, and 90\%, respectively. 
We use their optimal configuration to conduct comparison experiments.

\subsection{Topic Classification}
\label{B4}
About topic classification for adjusting semantic feature within mini-batch construction, we use the fastText model~\cite{joulin2016bag,joulin2016fasttext}. 
Specifically, we use the models trained on the DBpedia Ontology and Yahoo Answers datasets and then apply to the MIRACL training set for topic classification.
DBpedia Ontology is a dataset for text classification with 14 fine-grained entity categories, while Yahoo Answers provides 10 coarse-grained general categories. 
However, these categorizations might not be directly applicable to the MIRACL training set.
Thus, we combine both label categorizations to restructure the topic classification scheme that can better cover the MIRACL training set. The categorization details are shown in Table~\ref{tab:fulu2}.
The final distribution of topic classification on the MIRACL is shown in Table~\ref{tab:fulu3}.

\section{Human Validation}
To ensure the quality of the identified false negatives from LLMs, we conduct a validation study on a subset including high, medium, and low resources.
For each language, we randomly sample 100 false negatives for human validation with three annotators~\cite{wang2024user}. The evaluation criteria are based on a three-level rating scheme (0/1/2), which denotes the relevance between the potential false negative and the given query.
The results are shown in Table \ref{tab:fulu4}. We can see that most of the false negatives identified by LLMs are considered irrelevant (annotated as 0) in the validation subset, with an agreement rate of 82.5\% measured by Fleiss' Kappa among three annotators, demonstrating the correctness of LLMs' identification to some degree.

\section{Addtional Experimental Results}
\label{C}
We present additional experimental results in this section. 
For the main comparison with existing multilingual retrieval methods, the results with the Recall@100 score are reported in Table \ref{tab:fulu5}, where we can see our method still shows strong performance by outperforming most of the baselines.
For the results comparison among hard negative mining methods, results with NDCG@10 and Recall@100 metrics are shown in
Table \ref{tab:fulu6} and Table \ref{tab:fulu7} with different backbone models. 
We observe a consistently better performance of our method compared with the others, which demonstrates the superior generalizability across different backbone models of our approach.

\begin{table*}[!t]
    \centering
    \renewcommand{\arraystretch}{1.3}
    \resizebox{\textwidth}{!}{%
    \begin{tabular}{@{}l l c c c c c c c c c c c c c c c c c c@{}}
    \toprule
    \textbf{Model} & \textbf{Method} & \textbf{Avg} & \textbf{ar} & \textbf{bn} & \textbf{en} & \textbf{es} & \textbf{fa} & \textbf{fi} & \textbf{fr} & \textbf{hi} & \textbf{id} & \textbf{ja} & \textbf{ko} & \textbf{ru} & \textbf{sw} & \textbf{te} & \textbf{th} & \textbf{zh}\\
    \midrule
    \multirow{6}{*}{\rotatebox[origin=c]{90}{\centering mBERT}}       & Naive Top-K & 53.6 & 63.8 & 62.9 & 45.4 & 44.7 & 49.7 & 60.3 & 50.4 & 41.2 & 39.9 & 56.5 & 49.5 & 48.3 & 59.2 & 72.1 & 63.4 & 50.4 \\
    & Top-K shifted by N & 53.9 & 64.1 & 63.1 & 45.7 & 44.8 & 50.1 & 60.5 & 50.7 & 41.5 & 40.4 & 56.8 & 49.6 & 48.8 & 59.6 & 72.4 & 63.6 & 50.6 \\
                & TopK-Abs & 54.1 & 64.3 & 63.4 & 46.1 & 44.6 & 50.5 & 59.9 & 51.2 & 42.3 & 40.5 & 56.9 & 49.4 & 49.2 & 60.1 & 72.6 & 63.9 & 51.2 \\
          & TopK-MarginPos & 54.5 & 63.7 & 63.8 & 46.5 & 45.2 & 50.4 & 61.2 & 52.1 & 43.1 & 41.2 & 57.1 & 50.1 & 49.6 & 60.6 & 71.9 & 62.8 & 52.6 \\
            & TopK-PercPos & 54.6 & 64.2 & 63.6 & 46.2 & 45.4 & 50.3 & 61.1 & 52.3 & 43.3 & 41.4 & 57.0 & 50.3 & 49.1 & 60.6 & 72.3 & 63.9 & 52.8 \\
     & Ours & \textbf{57.9} & \textbf{66.2} & \textbf{65.5} & \textbf{48.8} & \textbf{48.5} & \textbf{52.7} & \textbf{64.8} & \textbf{53.4} & \textbf{45.4} & \textbf{44.9} & \textbf{61.9} & \textbf{56.2} & \textbf{55.5} & \textbf{64.3} & \textbf{76.3} & \textbf{67.4} & \textbf{54.8} \\
    \midrule
    \multirow{6}{*}{\rotatebox[origin=c]{90}{\centering mDPR}}       & Naive Top-K & 61.0 & 72.9 & 70.4 & 47.2 & 50.8 & 55.2 & 70.9 & 54.3 & 46.4 & 44.3 & 65.3 & 59.4 & 59.3 & 67.7 & 80.6 & 68.2 & 62.9 \\
       & Top-K shifted by N & 61.8 & 73.1 & 71.3 & 48.4 & 51.4 & 56.8 & 70.6 & 56.3 & 47.8 & 45.2 & 64.2 & 61.3 & 60.2 & 68.6 & 81.4 & 68.2 & 63.2 \\
                & TopK-Abs & 62.1 & 73.3 & 70.9 & 47.9 & 50.6 & 57.5 & 70.7 & 57.8 & 49.0 & 45.9 & 64.5 & 60.9 & 61.0 & 69.7 & 81.5 & 68.0 & 64.1 \\
          & TopK-MarginPos & 62.4 & 72.7 & 71.6 & 48.8 & 51.8 & 57.3 & 71.1 & 57.2 & 48.4 & 45.9 & 65.4 & 62.1 & 60.8 & 69.4 & 81.8 & 69.2 & 63.8 \\
            & TopK-PercPos & 62.2 & 73.4 & 70.6 & 49.2 & 51.6 & 57.1 & 71.3 & 57.5 & 47.8 & 45.5 & 66.2 & 61.4 & 61.3 & 69.2 & 82.2 & 68.8 & 62.8 \\
     & Ours & \textbf{66.8} & \textbf{75.7} & \textbf{74.1} & \textbf{55.9} & \textbf{55.8} & \textbf{61.6} & \textbf{75.8} & \textbf{60.7} & \textbf{53.8} & \textbf{52.1} & \textbf{71.2} & \textbf{67.8} & \textbf{65.1} & \textbf{73.9} & \textbf{84.2} & \textbf{74.8} & \textbf{65.9} \\
    
    \midrule
    \multirow{6}{*}{\rotatebox[origin=c]{90}{\centering mE5}}       & Naive Top-K & 63.5 & 72.9 & 73.2 & 48.9 & 48.6 & 56.4 & 74.5 & 52.2 & 58.1 & 49.4 & 68.1 & 63.1 & 63.3 & 71.6 & 80.2 & 76.5 & 58.3 \\
       & Top-K shifted by N & 63.8 & 73.1 & 73.4 & 49.3 & 49.2 & 56.6 & 74.8 & 52.5 & 57.9 & 49.6 & 68.5 & 64.4 & 63.5 & 71.9 & 80.4 & 76.8 & 58.5 \\
                & TopK-Abs & 64.0 & 73.3 & 73.6 & 49.7 & 49.6 & 56.8 & 74.4 & 53.1 & 58.2 & 49.8 & 69.1 & 64.6 & 63.9 & 72.3 & 80.8 & 77.2 & 57.8 \\
          & TopK-MarginPos & 64.2 & 73.4 & 73.9 & 49.9 & 49.7 & 57.1 & 74.1 & 53.4 & 58.6 & 50.2 & 69.2 & 64.3 & 64.1 & 72.6 & 81.2 & 77.6 & 58.4 \\
            & TopK-PercPos & 64.4 & 73.6 & 74.1 & 50.1 & 49.6 & 57.3 & 74.6 & 53.6 & 58.8 & 50.1 & 69.4 & 64.4 & 64.3 & 72.8 & 81.3 & 77.1 & 58.8 \\
     & Ours & \textbf{67.4} & \textbf{77.2} & \textbf{76.7} & \textbf{52.1} & \textbf{52.4} & \textbf{59.8} & \textbf{77.6} & \textbf{56.2} & \textbf{61.7} & \textbf{53.4} & \textbf{71.5} & \textbf{68.3} & \textbf{67.2} & \textbf{75.4} & \textbf{84.8} & \textbf{80.9} & \textbf{62.9} \\
    
    \bottomrule
    \end{tabular}%
    }
    \caption{Comparison of different hard negative construction methods based on different backbone models for multilingual retrieval evaluated on the MIRACL dataset with nDCG@10 scores. \textbf{Bold} indicates the best result based on the corresponding backbone model.}
    \label{tab:fulu6}
\end{table*}

\begin{table*}[!t]
    \centering
    \renewcommand{\arraystretch}{1.3}
    \resizebox{\textwidth}{!}{%
    \begin{tabular}{@{}l l c c c c c c c c c c c c c c c c c c@{}}
    \toprule
    \textbf{Model} & \textbf{Method} & \textbf{Avg.} & \textbf{ar} & \textbf{bn} & \textbf{en} & \textbf{es} & \textbf{fa} & \textbf{fi} & \textbf{fr} & \textbf{hi} & \textbf{id} & \textbf{ja} & \textbf{ko} & \textbf{ru} & \textbf{sw} & \textbf{te} & \textbf{th} & \textbf{zh}\\
    \midrule
    \cmidrule(lr){1-19}
    \multirow{6}{*}{\rotatebox[origin=c]{90}{\centering mBERT}}        & Naive Top-K & 83.6  & 87.6  & 88.1  & 78.8  & 80.5  & 85.1  & 86.5  & 85.7  & 77.4  & 71.8  & 86.5  & 79.3  & 81.4  & 86.4  & 90.2  & 85.1  & 87.2  \\
       & Top-K shifted by N & 83.8  & 87.9  & 88.3  & 79.2  & 80.1  & 85.4  & 86.7  & 85.2  & 77.6  & 72.6  & 86.8  & 79.7  & 81.8  & 86.7  & 90.6  & 85.4  & 87.4  \\
                 & TopK-Abs & 83.9  & 88.1  & 88.5  & 79.4  & 79.8  & 85.7  & 85.4  & 86.1  & 78.4  & 72.8  & 86.4  & 79.4  & 81.9  & 86.8  & 90.8  & 85.7  & 87.8  \\
           & TopK-MarginPos & 84.0  & 87.5  & 88.9  & 78.7  & 80.5  & 84.5  & 87.4  & 86.6  & 79.1  & 73.2  & 87.4  &  80.2 & 81.4  & 87.1  & 89.6  & 84.2  & 88.1  \\
             & TopK-PercPos & 84.4  & 88.4  & 88.4  & 79.7  & 80.7  & 84.9  & 87.2  & 86.8  & 79.3  & 73.6  & 87.2  & 80.5  & 81.6  & 87.2  & 89.9  & 85.9  & 88.3  \\
      & Ours & \textbf{87.9}  & \textbf{91.2}  & \textbf{92.1}  & \textbf{83.4}  & \textbf{83.2}  & \textbf{87.8}  & \textbf{90.8}  & \textbf{90.9}  & \textbf{81.6}  & \textbf{77.6}  & \textbf{91.1}  & \textbf{84.3}  & \textbf{86.2}  & \textbf{90.6}  & \textbf{94.9}  & \textbf{89.2}  & \textbf{90.8}  \\
    \midrule

    \multirow{6}{*}{\rotatebox[origin=c]{90}{\centering mDPR}}       & Naive Top-K  & 89.0  & 94.5  & 93.4  & 81.9  & 86.1  & 90.7  & 91.7  & 92.5  & 82.0  & 74.7  & 90.3  & 90.7  & 87.7  & 90.9  & 92.8  & 89.0  & 95.4  \\
       & Top-K shifted by N & 89.8  & 95.1  & 94.2  & 83.2  & 86.8  & 91.2  & 91.2  & 93.2  & 84.2  & 75.6  & 91.4  & 90.1  & 89.6  & 91.8  & 93.8  & 90.8  & 94.6  \\
                & TopK-Abs  & 90.1  & 94.8  & 93.9  & 82.4  & 85.2  & 92.0  & 91.5  & 94.6  & 86.7  & 76.2  & 92.0  & 88.8  & 91.0  & 92.7  & 93.6  & 90.4  & 95.9  \\
           & TopK-MarginPos & 90.3  & 94.4  & 94.4  & 83.8  & 87.2  & 91.8  & 92.0  & 94.2  & 86.1  & 75.5  & 92.4  & 89.4  & 90.8  & 92.4  & 94.2  & 91.0  & 95.4  \\
            & TopK-PercPos  & 90.0  & 94.8  & 93.8  & 84.3  & 86.8  & 91.8  & 92.0  & 94.4  & 82.6  & 74.6  & 92.7  & 89.8  & 91.4  & 92.2  & 94.4  & 90.8  & 94.2  \\
      & Ours & \textbf{94.4}  & \textbf{96.8}  & \textbf{96.3}  & \textbf{90.1}  & \textbf{90.8}  & \textbf{93.9}  & \textbf{96.6}  & \textbf{96.2}  & \textbf{90.2}  & \textbf{86.8}  & \textbf{96.6}  & \textbf{94.8}  & \textbf{94.5}  & \textbf{94.8}  & \textbf{97.4}  & \textbf{96.6}  & \textbf{97.6}  \\
    \midrule
    
    \multirow{6}{*}{\rotatebox[origin=c]{90}{\centering mE5}}         & Naive Top-K & 91.9  & 94.9  & 95.3  & 83.4  & 82.9  & 90.8  & 95.8  & 90.5  & 91.2  & 86.4  & 95.2  & 91.3  & 92.4  & 93.9  & 96.2  & 95.8  & 94.9  \\
       & Top-K shifted by N & 92.1  & 95.1  & 95.5  & 83.6  & 83.4  & 90.6  & 95.6  & 90.7  & 90.7  & 86.6  & 95.4  & 91.7  & 92.6  & 94.1  & 96.3  & 95.9  & 95.1  \\
                 & TopK-Abs & 92.1  & 95.3  & 95.2  & 83.8  & 83.6  & 90.4  & 95.2  & 90.8  & 91.3  & 87.1  & 95.8  & 91.8  & 92.3  & 94.3  & 96.6  & 96.2  & 94.6  \\
             & TopK-MarginPos & 92.2  & 95.4  & 95.1  & 84.1  & 83.8  & 90.9  & 94.9  & 90.4  & 91.6  & 86.8  & 95.7  & 91.4  & 92.5  & 94.6  & 96.4  & 96.4  & 94.8  \\
          & TopK-PercPos  & 92.2  & 95.6  & 94.7  & 84.4  & 83.4  & 90.6  & 95.4  & 89.8  & 91.7  & 86.6  & 95.6  & 91.3  & 92.8  & 94.5  & 96.8  & 96.2  & 95.2  \\
     & Ours  & \textbf{94.8}  & \textbf{97.4}  & \textbf{98.4}  & \textbf{86.9}  & \textbf{88.6}  & \textbf{93.4}  & \textbf{97.9}  & \textbf{92.2}  & \textbf{93.4}  & \textbf{88.4}  & \textbf{97.3}  & \textbf{94.4}  & \textbf{95.4}  & \textbf{96.2}  & \textbf{99.3}  & \textbf{99.2}  & \textbf{98.6}  \\
    \midrule

    \multirow{6}{*}{\rotatebox[origin=c]{90}{\centering BGE}}        & Naive Top-K & 92.9  & 97.5  & 98.2  & 85.2  & 84.2  & 90.9  & 96.5  & 91.7  & 90.7  & 79.3  & 94.7  & 94.1  & 94.6  & 96.0  & 98.8  & 98.4  & 95.2  \\
       & Top-K shifted by N & 93.2  & 97.2  & 98.5  & 85.8  & 85.8  & 91.6  & 96.3  & 92.6  & 91.0  & 82.2  & 94.4  & 94.2  & 94.2  & 95.6  & 98.6  & 98.2  & 95.4  \\
                 & TopK-Abs & 93.2  & 97.5  & 98.2  & 84.8  & 86.0  & 91.2  & 96.5  & 93.0  & 90.8  & 82.6  & 94.2  & 94.6  & 94.4  & 96.2  & 98.2  & 98.0  & 95.3  \\
          & TopK-MarginPos  & 93.4  & 96.9  & 98.0  & 86.2  & 87.2  & 91.7  & 96.8  & 93.2  & 91.6  & 83.4  & 93.8  & 94.2  & 93.8  & 96.4  & 98.5  & 98.6  & 94.6  \\
            & TopK-PercPos  & 93.6  & 97.7  & 98.4  & 85.8  & 87.6  & 91.4  & 96.1  & 93.1  & 91.8  & 83.6  & 94.9  & 94.7  & 94.1  & 96.1  & 98.8  & 97.8  & 95.6  \\
   
                         & Ours & \textbf{95.9}  & \textbf{97.9}  & \textbf{98.9}  & \textbf{89.9}  & \textbf{90.6}  & \textbf{94.9}  & \textbf{98.1}  & \textbf{95.4}  & \textbf{94.3}  & \textbf{90.4}  & \textbf{98.1}  & \textbf{96.1}  & \textbf{96.3}  & \textbf{97.0}  & \textbf{99.3}  & \textbf{99.1}  & \textbf{97.3}  \\
    \bottomrule
    \end{tabular}%
    }
    \caption{Comparison of different hard negative construction methods based on different backbone models for multilingual retrieval evaluated on the MIRACL dataset with Recall@100 scores. \textbf{Bold} indicates the best result based on the corresponding backbone model.}
    \label{tab:fulu7}
\end{table*}

\begin{table}[!t]
\centering
\renewcommand{\arraystretch}{1.2}
\begin{tcolorbox}[colback=gray!10, colframe=black!60, boxrule=0.5pt, arc=4pt, width=\textwidth, title=\textbf{Prompt for selecting false negatives}]
\textbf{\# Task Review:}\\
Your task is to evaluate a Candidate Answer based on a given Question and Standard Answer. Use the following two evaluation criteria to guide your assessment: \\

\textbf{\# Evaluation Criteria}\\

\textbf{\#\# Information Accuracy}\\
(1) \textbf{Definition}: Assess whether the Candidate Answer contains factual inaccuracies or misleading information. If a Standard Answer is provided, base your judgment on both the Question and the Standard Answer. If the Standard Answer is empty, evaluate based solely on the Question. \\
(2) \textbf{Scoring Guidelines:} 
\begin{itemize}
    \item \textbf{0}: The Candidate Answer contains clear factual errors or significantly misrepresents the meaning.
    \item \textbf{1}: The Candidate Answer has minor inaccuracies, but the overall meaning is still mostly correct.
    \item \textbf{2}: The Candidate Answer is entirely accurate with no factual errors.
\end{itemize}
\textbf{\#\# Information Completeness}\\
(1) \textbf{Definition}: Evaluate how well the Candidate Answer addresses the key aspects of the Question.
(2) \textbf{Scoring Guidelines:}
\begin{itemize}
    \item \textbf{0}: Major aspects of the question are not addressed or key points are missing.
    \item \textbf{1}: Most key points are addressed, but some minor details are omitted.
    \item \textbf{2}: All major and minor points are fully addressed.
\end{itemize}
\textbf{\# Input:}\\
\textbf{Question:} \{Input Question\}\\
\textbf{Candidate Answer:} \{Input Candidate Answer\}\\
\textbf{Standard Answer:} \{Input Standard Answer\}\\
\textbf{\# Output:} \\ 
\{"Information Accuracy": \{0/1/2\}, "Information Completeness": \{0/1/2\}\}
\end{tcolorbox}
\caption{Prompt for LLMs to judge false negatives.}
\label{tab:fulu9}
\end{table}

\end{document}